\documentclass[a4paper,11pt,preprintnumbers]{article}
\pdfoutput=1 

\usepackage{jcappub} 
\usepackage{booktabs}
\usepackage[T1]{fontenc} 
\usepackage{color}
\usepackage{graphicx, multirow, soul, url, amsmath, amsfonts, amssymb, mathrsfs}
\usepackage[all]{hypcap}
\usepackage{bbm}
\usepackage{cancel}
\usepackage{braket}
\usepackage[normalem]{ulem}
\usepackage{slashed}
\usepackage[dvipsnames]{xcolor}
\usepackage{multicol, blindtext}
\usepackage[section]{placeins}
\usepackage[version=3]{mhchem}
\usepackage{caption}
\usepackage{lipsum}
\usepackage{hyperref}
\usepackage{float}
\usepackage{mathtools}
\usepackage{footnote}
\usepackage{overpic, bm, epsfig}
\usepackage[splitrule]{footmisc}
\usepackage{bbold}
\usepackage{amsbsy}
\usepackage[export]{adjustbox}
\usepackage{units}
\usepackage{commath}
\usepackage{subcaption}

\newcommand{\be}{\begin{equation}}
\newcommand{\ee}{\end{equation}}
\newcommand{\bea}{\begin{eqnarray}}
\newcommand{\eea}{\end{eqnarray}}
\newcommand{\ba}{\begin{aligned}}
\newcommand{\ea}{\end{aligned}}

\DeclareRobustCommand{\secref}[1]{Section~\ref{#1}}

\newcommand{\equaref}[1]{Eq.~(\ref{#1})}
\newcommand{\equasref}[2]{Eqs.~(\ref{#1})~and~(\ref{#2})}

\newcommand{\secrefs}[2]{Sections~\ref{#1}~and~\ref{#2}}

\newcommand{\IPPP}{Institute for Particle Physics Phenomenology, Department of Physics, Durham University, Durham, DH1 3LE, United Kingdom}
\title{\huge Metastable Strings and Gravitational Waves in One-Scale Models}
\author[a]{James Ingoldby\,,}
\author[a]{Valentin V. Khoze\,}
\author[a]{and Jessica Turner}

\affiliation[a]{\IPPP}

\emailAdd{james.a.ingoldby@durham.ac.uk, valya.khoze@durham.ac.uk, jessica.turner@durham.ac.uk}

\abstract{
Metastable cosmic strings provide a minimal and predictive origin for the stochastic gravitational-wave background reported by Pulsar Timing Array experiments. We analyse this possibility in electroweak-like dark sectors with a single-stage breaking $SU(2)\times U(1)\!\to\!U(1)$ driven by one Higgs field. In the regime with dark sector Higgs mass below the $Z'$ mass, and for sufficiently small $W'$ mass, the resulting $Z$-string is classically stable but undergoes quantum decay via nucleation of monopole--antimonopole pairs along the string. We compute the corresponding semiclassical bounce action in a \emph{thin-defect} approximation, treating both the string core and the monopole endpoints as localised defects whose sizes are small relative to their separation in the tunnelling configuration. This yields a decay rate per unit length that depends on the gauge couplings and the mass hierarchy. We delineate the parameter space in which single-scale dark-sector models reproduce the PTA signal, demonstrating the applicability of the thin-defect approximation throughout the phenomenologically favoured region and without invoking extended Higgs sectors or multi-stage symmetry breaking.

}

\keywords{Metastable Strings, Semiclassical Effects, Gravitational Waves}

\begin{document}
\maketitle

\def\thefootnote{\arabic{footnote}}
\setcounter{footnote}{0}

%%%%%%%%%%%%%%%%%%%%%%%%%%%%%%%%%%%%%%%%%%%%%%%%%%
\section{Introduction}
Gravitational wave (GW) astronomy has entered a new era with the recent discoveries by Pulsar Timing Array (PTA) collaborations. In 2023, NANOGrav~\cite{NANOGrav:2023gor,NANOGrav:2023icp,NANOGrav:2023hfp,NANOGrav:2023ctt,NANOGrav:2023hvm}, EPTA~\cite{EPTA:2023fyk,EPTA:2023sfo,EPTA:2023akd,EPTA:2023gyr,EPTA:2023xxk,EuropeanPulsarTimingArray:2023egv}, PPTA~\cite{Zic:2023gta,Reardon:2023zen,Reardon:2023gzh}, and CPTA~\cite{Xu:2023wog} all reported compelling evidence for a stochastic GW background in the nanohertz frequency range. The detection of quadrupolar spatial correlations consistent with the Hellings–Downs curve~\cite{Hellings:1983fr} strongly indicates that the signal is of gravitational origin. While the most anticipated source is a cosmic population of inspiralling supermassive black hole binaries~\cite{Haehnelt:1994wt,Rajagopal:1994zj,Jaffe:2002rt,Wyithe:2002ep,Sesana:2004sp,Sesana:2008mz,Burke-Spolaor:2018bvk,Middleton:2020asl}, the origin is not yet confirmed, and several cosmological explanations remain plausible. These include first-order phase transitions~\cite{NANOGrav:2021flc,Kamionkowski_1994,Grojean:2006bp,Huber:2008hg,Jaeckel_2016,Ellis:2018mja,Nakai:2020oit,Ratzinger:2020koh,
Ashoorioon:2022raz,Freese:2022qrl,Morgante:2022zvc,Bringmann:2023opz,Winkler:2024olr}, annihilating domain walls~\cite{Ferreira:2022zzo}, cosmic strings~\cite{Blasi:2020mfx,Ellis:2020ena,Buchmuller:2020lbh,Samanta:2020cdk,King:2020hyd,Lazarides:2021uxv,Buchmuller:2023aus,Chitose:2023dam}, axion dynamics~\cite{Ratzinger:2020koh,Wang:2022rjz}, and GWs induced by large scalar fluctuations~\cite{Vaskonen:2020lbd,Kohri:2020qqd,DeLuca:2020agl,Zhao:2022kvz,Dandoy:2023jot}. The PTA era therefore offers a unique opportunity to probe physics far beyond the reach of terrestrial experiments.

In this work we explore the possibility that the PTA signal originates from \emph{metastable} cosmic strings. Unlike their stable counterparts, metastable strings decay over time, leading to a distinct, time-evolving GW spectrum. Their lifetime is determined by a quantum tunnelling process in which a segment of string nucleates a monopole–antimonopole pair that severs it~\cite{Preskill:1992ck}. The decay rate is exponentially suppressed by a semiclassical bounce action, conventionally parametrised by a dimensionless quantity~$\kappa$, whose magnitude controls both the amplitude and spectral slope of the resulting GW background. Current PTA data already prefer $\kappa$ values within a relatively narrow range~\cite{NANOGrav:2023hvm,Buchmuller:2023aus,Antusch:2023zjk,Madge:2023dxc}, motivating microscopic models that can realise this parameter naturally.
We focus on a simple and predictive setting capable of generating such metastable strings: an electroweak-like dark-sector gauge theory with a single-stage symmetry breaking,
$SU(2)\times U(1)\ \to\ U(1),
$
which can be viewed as a minimal analogue of the Standard Model’s electroweak sector. In this theory, the resulting $Z$-strings are not topologically protected but can be dynamically long-lived in the near-semi-local regime, where the $SU(2)$ coupling is small compared to the $U(1)$ one. Their decay proceeds through monopole–antimonopole nucleation, in contrast to multi-stage models ($SU(2)\!\to\!U(1)\!\to\!\emptyset$) where monopoles are topologically embedded from the outset~\cite{Buchmuller:2019gfy,Buchmuller:2023aus,Shifman:2002yi,Chitose:2023dam,Chitose:2025cmt}.

Our goal is to determine whether this one-scale framework can reproduce the $\kappa$ values suggested by PTA observations. We compute the classical string tension by solving the Nielsen–Olesen equations, obtaining the dimensionless parameter~$\alpha_{\rm str}$, which quantifies the dependence of the string energy on the Higgs-to-$Z$ mass ratio. We also calculate~$\alpha_\infty$, a complementary parameter describing the energy of the delocalised field configuration associated with monopole-like defects at the ends of the string. The ratio~$\alpha_\infty/\alpha_{\rm str}$ directly governs the bounce action and hence the decay rate. This mapping connects the fundamental Lagrangian parameters, gauge couplings and mass ratios, to the observable~$\kappa$ parameter. We then delineate the regions of parameter space where the theory is self-consistent, the semiclassical approximation remains valid, and the resulting GW spectrum can account for the PTA signal.

This paper is structured as follows. In \secref{sec:eWlike}, we introduce our electroweak-like model, define the $Z$-string ansatz, and discuss the semi-local limit crucial for metastability. In \secref{sec:tension}, we calculate the dimensionless tension coefficients for both the delocalised flux and the localised string. In \secref{sec:bounce}, we derive the bounce action for the string's quantum decay via monopole nucleation and relate it to the decay parameter $\kappa$. In \secref{sec:results}, we present our main findings, comparing our model's parameter space to the PTA-favoured region for $\kappa$ and the region of classical stability. We conclude in \secref{sec:conclusions} with a discussion of our results, the robustness of our approximations, and a comparison to other multi-scale models in the literature.

\section{Electroweak-like Model}\label{sec:eWlike}
We now develop the framework used to model metastable cosmic strings as a possible source of the nanohertz gravitational-wave background. \secref{sec:lagrangian} introduces the Lagrangian of an electroweak-like $SU(2)\times U(1)$ gauge theory and describes its particle content and single-stage symmetry breaking. We then present the classical $Z$-string solutions (ansatz) that arise within this model in \secref{sec:ansatz}. Finally, \secref{sec:topology} discusses the non-topological nature of the strings and establishes the conditions for metastability in the semi-local limit, which will be crucial for determining their decay rate and gravitational-wave signature.
%%%%%%%%%%%%%%%%%%%%%%%%%%%%%%%%%%%%%%%%%%%%%%%%%%%
\subsection{Lagrangian, symmetry breaking and mass eigenstates}\label{sec:lagrangian}
We consider a model with a gauge sector analogous to the Standard Model's electroweak sector, based on the gauge group $G = SU(2)\times U(1)$. This model should be understood as a component of a Beyond the Standard Model (BSM) or dark sector, rather than the Standard Model itself. This theory contains a single Higgs doublet field, $\Phi$, which transforms as a fundamental under SU(2). When $\Phi$ acquires a vacuum expectation value (vev) $\langle\Phi\rangle=\eta(0,1)/\sqrt{2}$, the symmetry is spontaneously broken to a residual $U(1)_{\rm IR}$ subgroup, 
\begin{equation}
\label{eq:Gsingle}
G\equiv SU(2) \times U(1)\, \rightarrow\, U(1)_{\rm IR}\,.
\end{equation}
A key feature of our setup is that it involves only a single stage of symmetry breaking, governed by one vev~$\eta$, as shown in \equaref{eq:Gsingle}\footnote{This is different from multi-scale models of metastable strings, where the symmetry breaking pattern is governed by different vevs, that are often considered in the literature, 
    see e.g.~\cite{Buchmuller:2019gfy,Buchmuller:2023aus,Shifman:2002yi,Chitose:2023dam,Chitose:2025cmt}.}.
 The Lagrangian of this theory is
\begin{equation}
\label{eq:Lbos}
  \mathcal{L} =
     -\frac{1}{4} W_{\mu\nu}^{a} W^{\mu\nu a}
     -\frac{1}{4} B_{\mu\nu} B^{\mu\nu}
     + |D_{\mu} \Phi|^{2}
     - \lambda \left(\Phi^{\dagger} \Phi - \frac{\eta^{2}}{2}\right)^{2}\,,
\end{equation}
where Greek letters \(\mu, \nu = 0,1,2,3\) denote spacetime indices,  Latin letters \(a, b, c = 1,2,3\) label the indices of \(SU(2)\), $ W_{\mu\nu}$ and $B_{\mu\nu}$ are the field strength tensors of $SU(2)$ and $U(1)$ which are defined as
\begin{align}
  W_{\mu\nu}^{a} &= \partial_{\mu} W_{\nu}^{a} - \partial_{\nu} W_{\mu}^{a}
                   + g\, \epsilon^{abc} W_{\mu}^{b} W_{\nu}^{c}\,, \\
  B_{\mu\nu}     &= \partial_{\mu} B_{\nu} - \partial_{\nu} B_{\mu}\,,
\end{align}
with $W_\mu^{a}$ $(a=1,2,3)$ and $B_\mu$ the gauge fields of $SU(2)$ and $U(1)$, respectively, and
$g$ and $g'$ their gauge couplings which we assume are real throughout. The covariant derivative acting on the Higgs doublet is
\begin{equation}\label{eq:covdiffH}
  D_{\mu} \Phi = \left(
      \partial_{\mu} - \frac{i g}{2} \tau^{a} W_{\mu}^{a}
                     - \frac{i g'}{2} B_{\mu} \right) \Phi\,,
\end{equation}
where the hypercharge of this theory is set to $Y=1$ and \(\tau^a\) are the Pauli matrices. The parameter \(\eta\) sets the vacuum expectation value of the Higgs field, while \(\lambda\) controls the strength of the Higgs self-interaction. The Higgs doublet may be written in component form as
\begin{equation}
\label{eq:Phidef}
  \Phi(x)=\frac{1}{\sqrt2}
  \begin{pmatrix}
     \phi_{1}+i\phi_{2}\\
     \phi_{3}+i\phi_{4}
  \end{pmatrix},\qquad
  \Phi^{\dagger}\Phi= \frac{1}{2}(\phi_1^2+\phi_2^2+\phi_3^2+\phi_4^2)= \tfrac12\eta^{2}\,.
\end{equation}
After spontaneous symmetry breaking, 
the resulting neutral mass eigenstates are an analogue of the $Z$ boson and the photon:
\be
  Z_{\mu}= -\sin\theta_{w}\,B_{\mu}+\cos\theta_{w}\,W_{\mu}^{3}\,,
  \qquad
  A_{\mu}=  \cos\theta_{w}\,B_{\mu}+\sin\theta_{w}\,W_{\mu}^{3}\,,
\ee
with the weak mixing angle given by $\sin^{2}\theta_{w} = {g'^{2}}/\left({g^{2}+g'^{2}}\right)$. The charged gauge bosons are 
\be
W_\mu^{ \pm} =\frac{\left(W_\mu^1 \mp i W_\mu^2\right) }{\sqrt{2}}\,,
\ee
and the tree-level masses of the electroweak gauge bosons and of the Higgs boson are respectively
\be
  M_{W} = \frac{g \eta}{2}\,,\qquad
  M_{Z} = \frac{\eta}{2} \sqrt{g^{2}+g'^{2}}\,, \qquad
  M_\Phi= \sqrt{2 \lambda} \eta\,.
\ee
For convenience, we define the parameter 
$\beta\equiv M_{\Phi}^{2}/M_{Z}^{2}$, which we will use throughout this work.
%%%%%%%%%%%%%%%%%%%%%%%%%%%%%%%%
\subsection{String ansatz and boundary conditions}\label{sec:ansatz}
%%%%%%%%%%%%%%%%%%%%%%%%%%%%%%%%
The model described above admits classical solutions that are static and uniform in one spatial direction~\cite{Vachaspati:1992fi}.  They are characterised by a pair of field configurations, $Z^{\,\rm str}_\mu(x_1,x_2)$ and $\Phi^{\rm str}(x_1,x_2)$,  that have no time ($t$) and $z$ ($x_3$) dependence and correspond to a string sourced by the $Z$ and Higgs fields along the $x_3$-direction:
\begin{align}
  \label{eq:Z-comp} Z^{\,\rm str}_\mu \,&=\,-\, \frac{2}{\sqrt{g'^2+g^2}}\, \frac{\delta_\mu^i\epsilon_{ij}x^j}{r^2}\,  V(r)\,,
  \\
 \label{eq:Phi-comp} 
  \Phi^{{\rm str}} \,&=\, e^{in\varphi} \, \frac{\eta}{\sqrt{2}}\, P(r) \, \begin{pmatrix}0\\ 1 \end{pmatrix}\,,
\end{align}
where $i,j=1,2$, the radial and azimuthal coordinates in the $x_1-x_2$-plane, $r^2=x_1^2+x_2^2$, $\varphi\equiv\arg(x_1+i x_2)$
and $V(r)$ and $P(r)$ are the string profile functions that are determined numerically by solving the classical equations of motion subject to the finite-energy boundary conditions,
\be
\label{eq:bc-s}
P(0)\,=\,0\,=\,V(0)\,, \qquad
P(\infty)\,=\,1\,=\,V(\infty)\,,
\ee
which ensure regularity at the string core and approach to the vacuum at infinity, yielding quantised flux. The remaining gauge fields vanish:
\begin{align}
 W^{1\,{\rm str}}_\mu \,&=\, W^{2\,{\rm str}}_\mu\,=\,
  A^{\,{\rm str}}_\mu \,=\, 0\,.
  \label{eq:W-comp} 
 \end{align}
The field configurations of \equasref{eq:Z-comp}{eq:Phi-comp} are known as the electroweak strings, or $Z$-strings, and amount to a non-Abelian embedding of the Abrikosov-Nielsen-Olesen (ANO)  classical string of the Abelian Higgs model~\cite{Abrikosov:1956sx,Nielsen:1973cs}. The index $n \in \mathbb{Z}$ is the magnetic flux of the $Z$-string, and in what follows, we will restrict to an elementary string with $n=1$. Generally, the field configurations described by the ansatz of \equasref{eq:Z-comp}{eq:Phi-comp} are not guaranteed or expected to be stable throughout the model’s parameter space. 

\subsection{Topology and the semi-local limit}\label{sec:topology}
Unlike the ANO vortex of the Abelian Higgs model, which is a \emph{topological} string protected by the non-trivial first homotopy group \(\pi_{1}\bigl(U(1)\bigr)=\mathbb Z\), the \(Z\)-string is \emph{not} topological.  After the symmetry breaking
the vacuum manifold is simply connected, such that 
\be
\pi_{1}\!\left({\left( SU(2)\times U(1)\right)}/{U(1)_{\rm IR}}\right)=0\,,
\ee
and no topologically stable vortex solutions exist.
Instead, \(Z\)-strings are \emph{non-topological solitons}, $i.e.$, finite energy classical field configurations whose (meta)stability is determined dynamically rather than by topology. 

Although not protected by topology, Refs.~\cite{Hindmarsh:1991jq,James:1992wb} showed that such electroweak-like string configurations can be classically stable local minima\footnote{A note on terminology: A field configuration is called classically stable if it is a local minimum of the energy functional $E$. In particular, there should be no negative or zero modes when expanding $E$ around a classically stable field configuration. There can, and in general will exist other local minima with lower values of $E$ separated by finite or infinite energy barriers from the classical solution. When the barrier is finite, the classical string can decay via quantum tunnelling into the true minimum. In this case the classical string is metastable and, in favourable conditions, its quantum decay rate can be computed semiclassically~\cite{Preskill:1992ck} using a version of Coleman's false vacuum decay approach~\cite{Coleman:1977py}.} of the energy per unit length in a restricted region of parameter space when the model becomes semi-local:
\be
\label{eq:semiloc}
g/g' \,\to\, 0\,,
\ee
 and \(M_{\Phi}<M_Z\) or equivalently $\beta < 1$.
Physically, classical stability of the string configuration of Eqs.~\eqref{eq:Z-comp} and \eqref{eq:Phi-comp} arises from two effects. First, when the Higgs is lighter than the $Z$ boson, its field provides an energetically favourable barrier that confines the $Z$-flux, creating a local energy minimum. Second, the $g \ll g'$ condition suppresses decays into $W^{\pm}$~bosons; in this limit ($\theta_w \to \pi/2$), the $Z$ boson aligns with the $U(1)_Y$ gauge boson ($Z_\mu \approx -B_\mu$), causing the $Z$-string to decouple from the purely $SU(2)$ sector as their interaction coupling $g \to 0$.

We demonstrate this stability in the semi-local limit, where the \(SU(2)\) sector of the gauge group \(G\) becomes a global symmetry, while the \(U(1)\) factor remains gauged. 
The resulting semi-local model then possesses a topologically conserved, integer-valued magnetic flux~\cite{Vachaspati:1991dz,Preskill:1992bf} (see also sec.~IID.2 in~Ref.~\cite{Preskill:1992ck} for a concise summary of the topological argument),
which corresponds to the quantum number \(n \in \mathbb{Z}\) appearing in the string field configurations~\equasref{eq:Z-comp}{eq:Phi-comp}. As mentioned, in what follows, we will restrict to the elementary flux sector with \(n=1\).
Stability is determined by an energy comparison between two configurations: the localised $Z$-string, where energy is confined in a flux tube of tension \(\mu_{\rm str}\), and a delocalised texture, where the same flux spreads across a broad scalar field configuration of tension \(\mu_\infty\):
\begin{equation}\label{eq:alpsdef}
    \mu_{\rm str}=\eta^2\,\alpha_{\rm str}(\beta),\qquad
\mu_\infty=\eta^2\,\alpha_\infty\,,
\end{equation}
where  \(\alpha_\infty\) and \(\alpha_{\rm str}\) are dimensionless quantities. 
We will compute them in \secrefs{sec:calcalphainf}{sec:calcalphas} respectively.
Physically, $\alpha_{\rm str}$ is a dimensionless quantity that parametrises the energy cost to confine one unit of the $Z$-flux in a localised string core, while $\alpha_\infty$ parametrises the energy density contained within a delocalised `spread' configuration.
The string, once formed, decays via quantum nucleation of field configurations that locally resemble the global monopole-antimonopole pairs. In the thin-defect limit, the barrier height is controlled by the Coleman's bounce action,
\begin{equation}
S_B \;=\; \frac{8\pi^2}{g^2} \, \frac{\alpha_\infty}{\alpha_{\rm str}}\,,
\end{equation}
which we compute in \secref{sec:bounce}. 

\section{Tension Parameters and Classical Stability}\label{sec:tension}
This section computes the two dimensionless tension coefficients that control the $Z$-string’s stability and decay rate. In \secref{sec:calcalphainf} we derive $\alpha_\infty$ (delocalised flux, sigma-model limit); in \secref{sec:calcalphas} we compute the string tension $\alpha_{\rm str}(\beta)$ numerically as a function of the mass ratio $\beta$.

\subsection{Computing the $\alpha_\infty$ tension coefficient}\label{sec:calcalphainf}

To compute \(\alpha_{\infty}\), we evaluate the energy of static two-dimensional configurations carrying one unit of magnetic flux, uniformly delocalised over a disc of radius \(R\) and taking the $R\to \infty$ limit.
Following  Ref.~\cite{Preskill:1992ck} (sec.~VIIIA), we can make two simplifying assumptions. For these infinitely-spread finite energy configurations, the Higgs field takes values in the vacuum manifold everywhere, 
\be
\label{eq:Phi-VM}
|\Phi_{\rm cl}(x)|^2=\eta^2 / 2\,,
\ee
and, consequently,  there are no contributions from the Higgs potential energy. Secondly, for the configuration size $R$ being large, the Coulomb energy density of the magnetic flux ${\cal E}_{\rm mg} \propto (\vec B)^2 \propto 1/R^4$, can also be neglected, and the only contribution to the energy comes from the gradients of the scalar fields. We can thus reduce our model to the one described by the Lagrangian, 
\be
\label{eq:sigmU1}
{\cal L}_{U(1)}\,=\, |D_\mu \Phi|^2
\,+\,
\lambda_{\rm L.m.}\left(|\Phi(x)|^2-\eta^2/2\right)\,,
\ee
Here $\lambda_{\rm L.m.}$ is the Lagrange multiplier, with mass dimension two, which enforces the vacuum manifold constraint of \equaref{eq:Phi-VM}. 
We work in the semi-local limit $g / g^{\prime} \rightarrow 0$, in which the non-Abelian gauge bosons decouple and the covariant derivative $D_\mu \Phi$ involves only the Abelian $U(1)$ gauge fields.

The model of \equaref{eq:sigmU1} is gauge-invariant which implies a redundancy in the number of field degrees of freedom. Only different gauge orbits should be counted as independent fields, while all fields in the same $U(1)$ gauge orbit are identified,
\be
\label{eq:orbit}
\Phi_\alpha (x) \,\equiv \, e^{i \xi(x)}\, \Phi_\alpha (x)\,.
\ee
Here $\Phi_{\alpha=1,2}$ denotes the complex Higgs doublet field and $e^{i \xi(x)}$ is the gauge transformation. Since the magnitude of the Higgs field in~\equaref{eq:sigmU1} is automatically fixed by the constraint~\equaref{eq:Phi-VM}, we can promote the field identification 
condition~\equaref{eq:orbit} to  $\Phi_\alpha (x) \,\equiv \, z(x)\, \Phi_\alpha (x)$ where $z(x)$ is an arbitrary complex function. This is equivalent to saying that the $U(1)$ gauge theory~\equaref{eq:sigmU1} has an equivalent description in terms of a  
$CP^1$ non-linear sigma model,
\be
\label{eq:sigmCP1}
{\cal L}_{CP^1}\,=\, \partial_\mu \Phi^\dagger_\alpha \, g(\Phi)_{\alpha \beta}\, 
\partial^\mu \Phi_\beta
\,+\,
\lambda_{\rm L.m.}\left(|\Phi(x)|^2-\eta^2/2\right)\,,
\ee
where $g(\Phi)_{\alpha \beta}$ is the metric tensor of the $CP^1$ theory
in affine coordinates, given by the Fubini-Study metric, see
$e.g.$ \cite{Eguchi:1980jx,Hindmarsh:1992yy},
\be
g(\Phi)_{\alpha \beta}\,=\, \delta_{\alpha \beta} \,-\, \frac{\Phi_\alpha \Phi^\dagger_\beta}{|\Phi|^2}\,.
\ee 
With a judicious change of variables,
\begin{equation}
\label{eq:phiaVM}
    \varphi^a \;=\; \frac{\Phi^\dagger \sigma^a \Phi}{\eta/\sqrt{2}}\,,
\end{equation}
we can linearise the 
$CP^1$ theory kinetic term, by noticing that
\be
\label{eq:expr}
\frac{1}{4}\, \partial_\mu \varphi^a \partial^\mu \varphi^a\,=\,
\partial_\mu \Phi^\dagger \cdot \partial^\mu \Phi \,-\,
\frac{(\partial_\mu \Phi^\dagger \cdot \Phi)(\Phi^\dagger \cdot \partial^\mu \Phi)}{\eta^2/2}\,=\, 
\partial_\mu \Phi^\dagger_\alpha \, g(\Phi)_{\alpha \beta}\, 
\partial^\mu \Phi_\beta\,.
\ee
In deriving the expression above we have used the Fierz identity for Pauli matrices, \newline $(A\sigma^aB)(C\sigma^aD)=2(A\cdot D)(C\cdot B)-(A\cdot B)(C\cdot D)$ and the constraint on the absolute value of the Higgs doublet $\Phi^\dagger \cdot \Phi = \eta^2/2$ in \equaref{eq:Phi-VM}. Consequently, we find that the first term on the RHS of \equaref{eq:sigmCP1} is simply the standard flat-space kinetic term $\frac{1}{4}\, \partial_\mu \varphi^a \partial^\mu \varphi^a$ with the normalisation factor $1/4$.

The real-valued fields $\varphi^a$ defined in~\equaref{eq:phiaVM} satisfy the relation $(\varphi^1)^2 +(\varphi^2)^2 +(\varphi^3)^2 =\eta^2/2$ and  conveniently parameterise the vacuum manifold of our original Abelian Higgs model,
\be
\frac{SU(2)\times U(1)}{U(1)_{\mathrm{IR}}}\;\simeq\;S^{2}\,.
\label{eq:VacM}
\ee
Thanks to~\equaref{eq:expr}, the
energy functional obtained from~\equaref{eq:sigmCP1} reduces to that of the
\((2+1)\)-dimensional nonlinear \(O(3)\) sigma model,
constructed solely from ordinary derivatives of \(\varphi^a\):
\be
\label{eq:sigm}
E\,=\, \int d^2 x\left(
\frac{1}{4}\, \left(\vec\partial \varphi^a \right)^2
\,+\,
\lambda_{\rm L.m.}((\varphi^a)^2-\eta^2/2)
\right)\,.
\ee
Here  \(E\) is the transverse energy, $i.e.$ the $3{+}1$D string tension (energy per unit length).
The model of \equaref{eq:sigm} has topological solitons~\cite{Polyakov:1975yp,Rubakov:2002fi}
with topological charges $n \in \pi_2(S^2)\sim \mathbb{Z}$. 
The single soliton solution is given by,
\be
\label{eq:soliton}
\varphi_{\rm cl}^{1,2} \,=\, \frac{\eta}{\sqrt{2}}\,  x_{1,2}\,\, \frac{2r_0}{r_0^2 +r^2}\,,
\qquad \varphi_{\rm cl}^{3} \,=\, \frac{\eta}{\sqrt{2}}\,\,\frac{r_0^2 - r^2}{r_0^2 +r^2}\,,
\ee
where $r^2=x_1^2 +x_2^2$ and $r_0$ is an arbitrary constant that plays the role of the soliton size.
The expressions shown in ~\equaref{eq:soliton} can be checked explicitly by verifying that they satisfy equations of motion and give a local minimum of the energy functional with 
\be
\label{eq:solE}
E^{\,\rm cl} \, =\, \pi\, \eta^2\,.
\ee
More detail and a derivation of the soliton solution above can also be found in Sec.~7 of the Ref.~\cite{Rubakov:2002fi}.  
In the sigma model of \equaref{eq:sigm}, the soliton size parameter $r_0$ is a flat direction. But the sigma model soliton provides an accurate approximation to the large-size solution of our $SU(2)\times U(1)$ semi-local model only for large scale sizes. Hence, we can identify $r_0$ with $R$ when we take the latter to infinity. By equating the two-dimensional soliton mass in ~\equaref{eq:solE} with the string tension $\mu_\infty$ of the infinite-size magnetic flux configuration of ~\equaref{eq:alpsdef}, we conclude that
\begin{equation}
    \label{eq:alphainf_final}
    \alpha_\infty \;=\; \pi \,.
\end{equation}

\subsection{Computing the string tension coefficient}\label{sec:calcalphas}

The dimensionless quantity $\alpha_{\rm str}$ is determined by the energy per unit length of the metastable string (tension) $\mu_\text{str}$, as indicated in Eq.~(\ref{eq:alpsdef}). This tension is obtained by solving the Nielsen--Olesen equations for the radial profile functions $P(\tilde{r})$ and $V(\tilde{r})$, which describe, respectively, the magnitude of the Higgs field and the azimuthal component of the $Z^0$ gauge field. These functions were introduced in Eqs.~(\ref{eq:Z-comp}) and (\ref{eq:Phi-comp}) and are solutions to
\begin{gather}
    P'' + \frac{P'}{\tilde{r}} - \frac{(1 - V)^2}{\tilde{r}^2} P + \beta (1 - P^2) P = 0 \,, \label{eq:NOP}\\
    V'' - \frac{V'}{\tilde{r}} + 2 (1 - V) P^2 = 0 \,, \label{eq:NOV}
\end{gather}
where $\tilde{r} = r M_Z/\sqrt{2}$ is the radial distance in units of the $Z^0$ boson mass and $\beta = M_\Phi^2 / M_Z^2$ parametrises the Higgs mass in the same units. For a given value of $\beta$, these coupled differential equations are solved numerically with boundary conditions from Eq.~(\ref{eq:bc-s}), ensuring regularity at the origin and vacuum behaviour at large $\tilde{r}$.

\begin{figure}[t!] 
    \centering
    \begin{subfigure}[t]{0.5\linewidth}
        \centering
        \includegraphics[width=\linewidth]{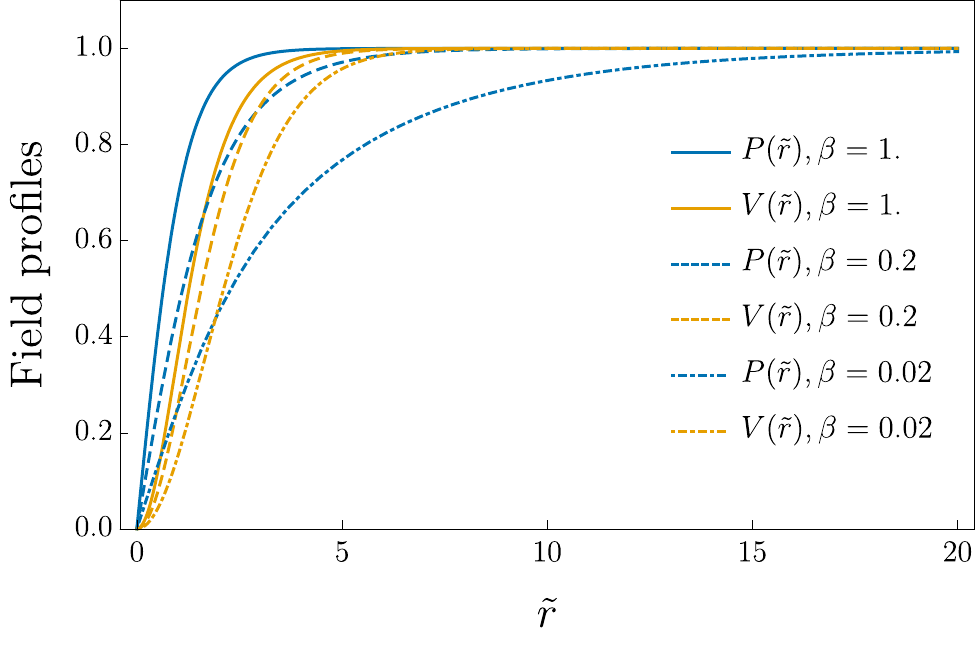}
        \caption*{}
    \end{subfigure}\hfill
    \begin{subfigure}[t]{0.5\linewidth}
        \centering
        \includegraphics[width=\linewidth]{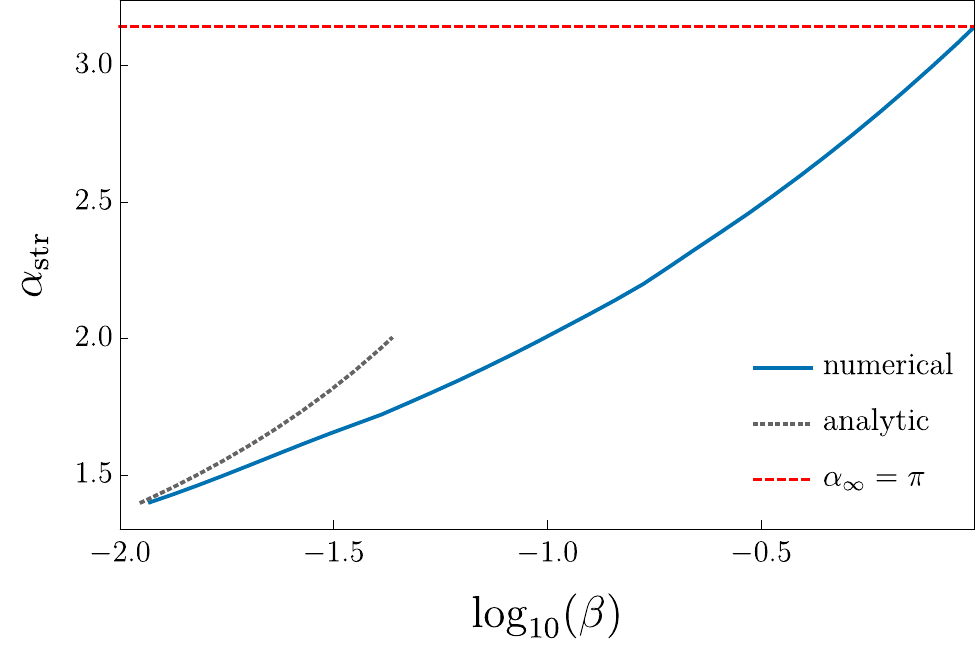}
        \caption*{}
    \end{subfigure}
    \caption{Left: Profile functions $P(\tilde{r})$ and $V(\tilde{r})$ for the classical string configuration. The functions are plotted for $\beta=0.02,\,0.2$ and 1. These functions are independent of $g$.
    Right: The blue line shows the dependence of $\alpha_\text{str}$ on $\log\beta$, for the range $-2<\log_{10}\beta<0$, where the string solution can be shown to be metastable. Note also that $\alpha_\text{str}$ depends on $\beta$ only, and has no further dependence on $g$. The grey dotted line shows the analytical approximation for $\alpha_{\rm str}$ valid in the small-$\beta$ limit and the red line $\alpha_{\rm str}=\pi$.}
     \label{fig:profilefuncs}
\end{figure}

Representative solutions are shown in the left panel of \figref{fig:profilefuncs} for three illustrative values of $\beta$.  
For small $\beta$ the scalar field recovers its vacuum value only slowly, while larger $\beta$ leads to a more localised Higgs core.  
The gauge profile $V(\tilde{r})$ adjusts accordingly so that the magnetic flux remains quantised.  
These profiles are independent of the gauge coupling $g$ once expressed in terms of the rescaled radius $\tilde{r}$.

The energy per unit length of the string is obtained by substituting the solutions of Eqs.~(\ref{eq:NOP}) and (\ref{eq:NOV}) into the static energy functional,
\begin{align}
    \mu_{\text{str}}
    = \pi \eta^2 \int_0^\infty \! \tilde{r} \, d\tilde{r} \,
    \bigg[ \left(P'\right)^2
          + \frac{1}{2\tilde{r}^2}\left(V'\right)^2
          + \frac{(1 - V)^2}{\tilde{r}^2} P^2
          + \frac{\beta}{2}\left(P^2 - 1\right)^2
    \bigg]\,.
    \label{eq:stringtension}
\end{align}
The dependence of $\alpha_{\rm str}$ on the mass ratio $\beta$ is displayed in the right panel of \figref{fig:profilefuncs}.  
As expected from the behaviour of the profile functions, $\alpha_{\rm str}$ increases slowly as $\beta$ increases: a heavier Higgs field localises the scalar core and slightly raises the overall tension. The role of the $SU(2)$ gauge coupling $g$ is implicit in the choice of dimensionless variables and does not appear in the energy functional~\cite{Vachaspati:1992fi}, or alter the shape of the curves shown in the right panel of \figref{fig:profilefuncs}, although it re-enters when converting back to physical units.
The string tension coefficient $\alpha_{\rm str}(\beta)$ as the function of the $\beta$ parameter is computed numerically in our model and plotted in the right panel of \figref{fig:profilefuncs}. 
This is the main result of this section along with the result for $\alpha_\infty$ in \equaref{eq:alphainf_final}. In the following section these quantities will be used to determine the semiclassical decay rate of the metastable string.

We observe that for $\beta = 1$ the string coupling satisfies $\alpha_{\rm str} = \pi$, coinciding with the value for $\alpha_\infty$ and saturating the Bogomolnyi bound~\cite{Bogomolny:1975de} for the energy. In addition, in the $\beta \to 0$ limit the ANO string tension has been obtained analytically in Ref.~\cite{Yung:1999du},
 \be
\lim_{\beta \to 0}\, \alpha_{\rm str} \,=\,
\lim_{\beta \to 0}\,  \frac{\mu_{\rm str}}{\eta^2} \,=\, \frac{2\pi}{\log(1/\beta)} \,.
\ee
This small-$\beta$ behaviour is in agreement with the corresponding asymptotic form of $\alpha_{\rm str}(\beta)$ extracted from our plot in~\figref{fig:profilefuncs}. 

\section{Bounce Calculation}\label{sec:bounce}
\begin{figure}[t!]
\centering
\includegraphics[width=0.45\textwidth]{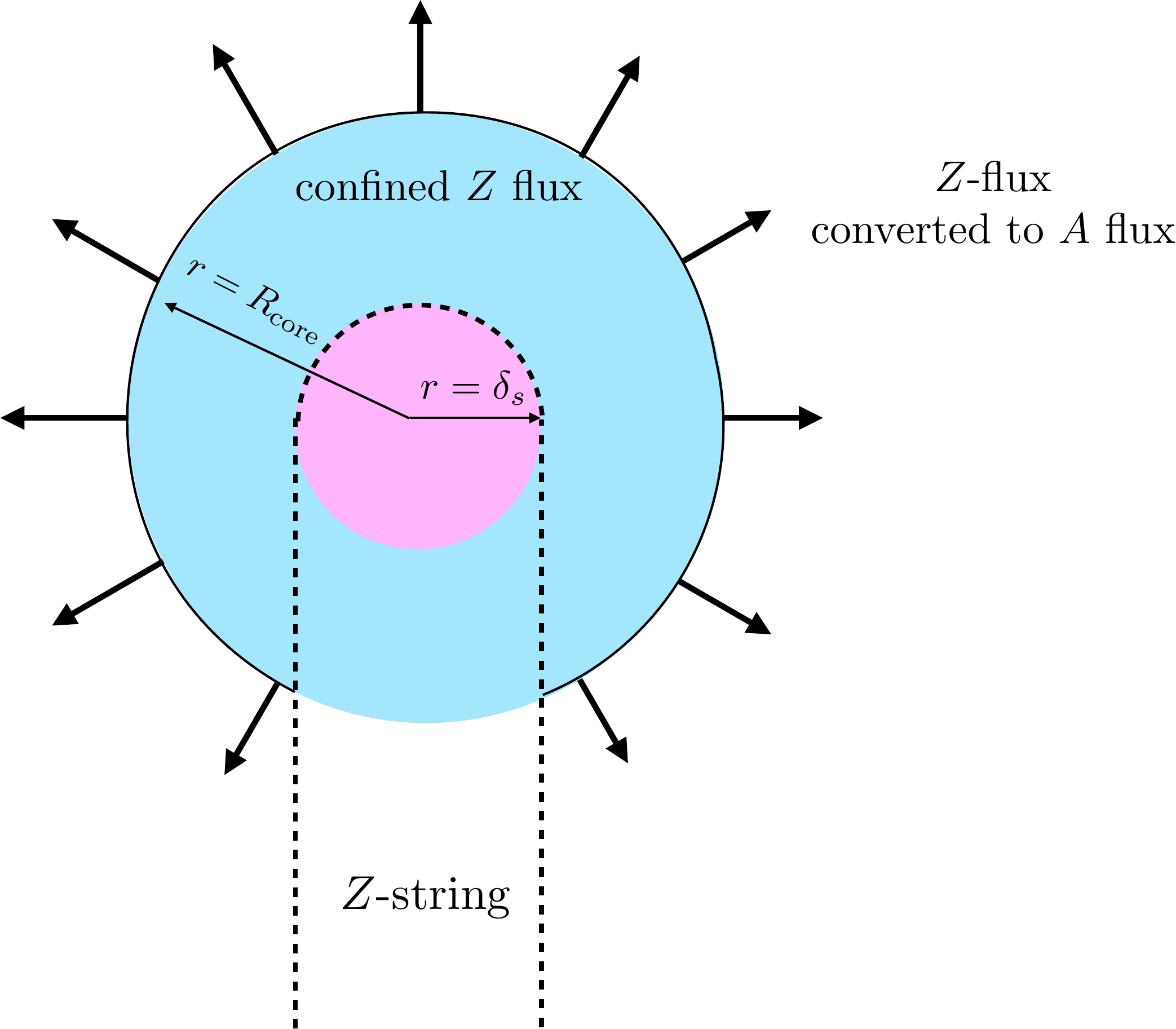}\hfill
\includegraphics[width=0.4\textwidth]{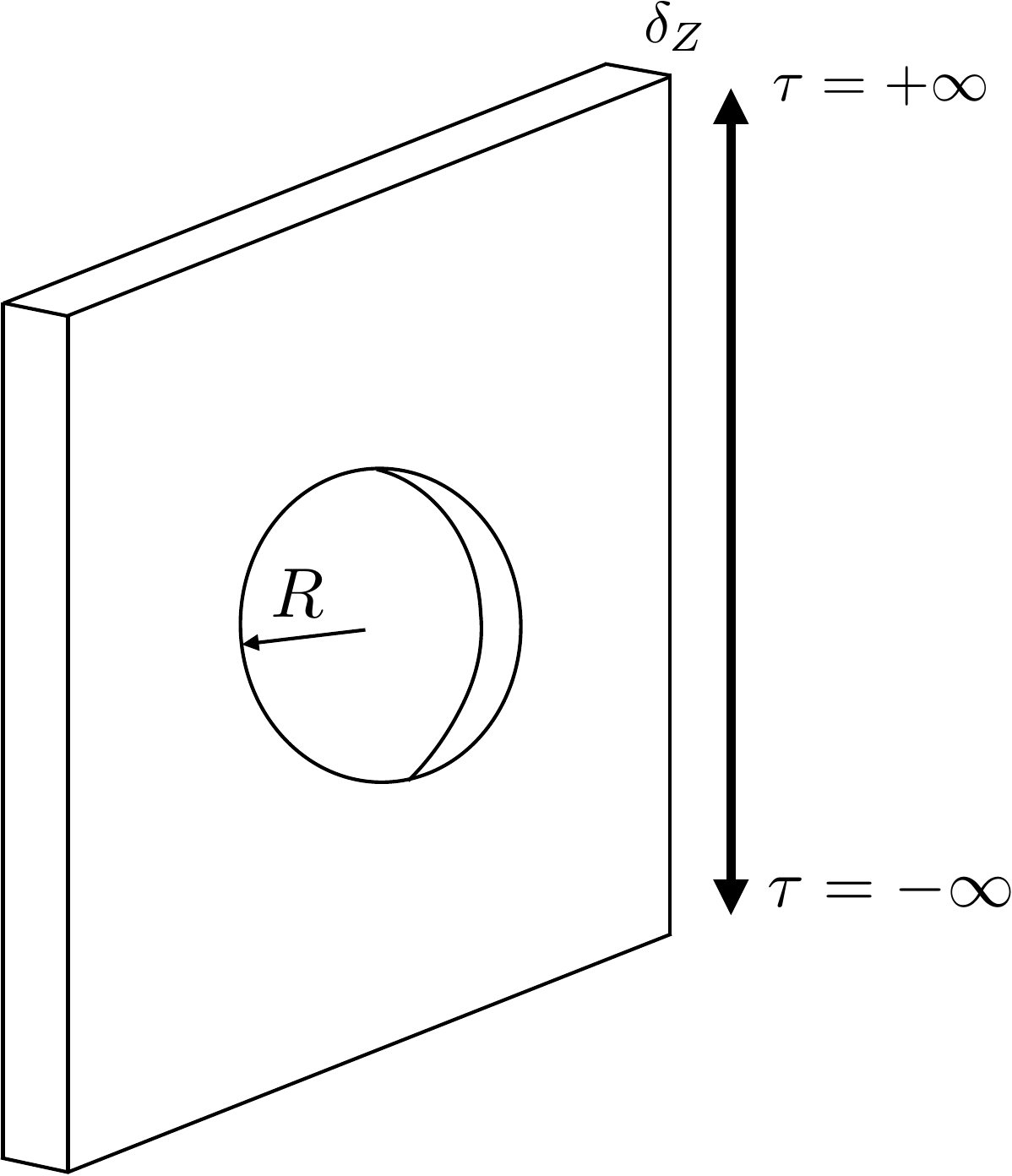}
\caption{Left: $Z$-string ending on a Nambu monopole with core size $R_{\rm core}$. Right: Euclidean worldsheet of a $Z$-string (width $\delta_s$) with nucleation of a monopole–antimonopole pair, whose closed worldline appears as a circle.}\label{fig:zstringnucl}
\end{figure}

In the semi-local limit where the $SU(2)$ symmetry is global, the string configuration is stable since the magnetic flux carried by the string is topological and hence conserved, and with  \(\alpha_{\rm str} < \alpha_{\infty}\), the localised string configuration is energetically more favourable than the delocalised lump with magnetic flux. 

When we turn on $g\neq 0$ the $SU(2)$ group is gauged and the string magnetic flux is no longer protected by topology as it can `unwind' by exciting non-Abelian gauge field degrees of freedom. The string is no longer absolutely stable but there is a region in the parameter space where it is still a local minimum now separated by a \emph{finite} energy barrier from minima with lower energy. This classical stability region of the string was established in Ref.~\cite{James:1992wb}, and is shown in \figref{fig:result} as the area to the right of the black contour on the $(\beta, \sin^2 \theta_w)$ plane. 
The classical string decays by quantum tunnelling and
in this section, we will derive the bounce action which determines the semiclassical exponent of the string decay rate following the thin-defect aka thin-wall approach of~\cite{Preskill:1992ck}.

To evaluate the tunnelling trajectory, we begin by observing that since the string is not topological, it can terminate on a source carrying magnetic field of the Abelian $A_\mu$.
The string configuration involves both the Higgs and \(Z\)-fields, which form string-like solutions as given in~\equaref{eq:Z-comp} and~\equaref{eq:Phi-comp}.
Inside the string core, there exists a non-vanishing magnetic \(Z\)-flux: 
\begin{equation}
\Phi_Z = \frac{4 \pi \sin \theta_w}{g'}\,,
\end{equation}
which can be obtained from integrating the \(Z\)-associated magnetic field over a surface in the \(x_1\)–\(x_2\) plane. Physically, this flux is confined within the core of the $Z$-string and is sourced by the winding of the Higgs field. At the endpoint of the string, the non-Abelian magnetic flux associated with the $W_\mu^3$ component of 
\be
  Z_{\mu}= -\sin\theta_{w}\,B_{\mu}+\cos\theta_{w}\,W_{\mu}^{3}\,,
\ee  
can terminate, while the magnetic flux of the $B_\mu$ component is converted to the magnetic flux of the massless photon field $A_\mu$. The end points of the $Z$-string are electromagnetic monopoles of Nambu~\cite{Nambu:1977ag,Kibble:2015twa},\footnote{These configurations exist only as string endpoints, i.e. they are always confined by the string and do not appear as free classical solutions. In practice they are 't Hooft--Polyakov-type monopole configurations composed of the $W_\mu^a$ gauge field and the adjoint Higgs field $n^a= \Phi^\dagger \tau^a \Phi/(\Phi^\dagger \Phi)$. This composite scalar field configuration is not invertible for $\Phi(x)=0$, which is precisely the point where the $Z$-string is attached.}
they absorb the string's \(Z\)-flux and continuously convert it into \(A\)-flux,
\be
\Phi_A = \frac{4 \pi \sin \theta_w }{g}\,,
\ee
associated with the massless photon.
In the limit \(g \ll g'\), or equivalently \(\sin\theta_w \approx 1\), the thickness of the string core is \(\delta_s \sim m_Z^{-1}\), and the  monopole has a larger characteristic size, \(R_{\text{core}} \gg \delta_s\).  Within the string core, the magnetic flux is carried by the massive \(Z\)-field and remains confined up to a characteristic distance \(\delta_s \). Beyond this region, at a radius \(R_{\text{core}} \gg \delta_s\), the flux is smoothly converted into the massless \(A\)-field, resulting in a long-range photon flux spreading spherically from the monopole. This configuration is schematically shown in the left diagram of \figref{fig:zstringnucl}.
The total core energy receives three contributions: the gradient energy of the Higgs field, which grows linearly with \(R_{\text{core}}\); the magnetic self-energy of the confined \(Z\)-flux, which vanishes for large \(R_{\text{core}}\) and the Coulomb energy of the \(A\)-flux, which scales inversely with \(R_{\text{core}}\). 
The energy contained in the $Z$ and $A$ field is obtained from integrating the magnetic field contained within the relevant volume:
\be
E=\frac{1}{2} \int_{r_1}^{r_2} {\cal B}^2 \, d^3 V\,,
\ee
and considering the hierarchy of scales shown in the left panel of  \figref{fig:zstringnucl}.
Summing these three contributions gives the total energy contained within the monopole core~\cite{Preskill:1992ck}:
\begin{equation}\label{eq:core}
E_{\text{core}} \sim \alpha_{\infty} \eta^2 R_{\text{core}} + \frac{2\pi \sin^2\theta_w}{g'^2}\left(m_Z - \frac{1}{R_{\text{core}}} \right) + \frac{2\pi \sin^2\theta_w}{g^2} \frac{1}{R_{\text{core}}} \,.
\end{equation}
Minimising this energy with respect to \(R_{\text{core}}\) gives
\be
\label{eq:Rmono}
R_{\rm core}
= \frac{1}{g\eta}\,\sqrt{\frac{2\pi\left(g'^2-g^2\right)}{\alpha_{\infty}\!\left(g^2+g'^2\right)}}
\approx
\sqrt{2} \frac{1}{g \eta}\,,
\ee
where we expanded in $g^2/g'^2\ll 1$ and substituted $\alpha_\infty=\pi$ in the second step.
We note that $R_{\rm core} \sim (g\eta)^{-1}$ is of order the $W$-boson Compton wavelength in this model, meaning the monopole’s core size is set by the heavy gauge boson scale. Substituting back into \equaref{eq:core} gives
\begin{equation}\label{eq:Ecore}
\begin{aligned}
E_{\text{core}}(R_{\rm core}) &\approx \sqrt{8\pi \alpha_{\infty}}\, \frac{\eta}{g} \,,
\end{aligned}
\end{equation}
where again we expanded in $g^2/g'^2\ll 1$\footnote{We note that in \cite{Preskill:1992ck}, there is an error in the calculation of $R_{\rm core}^2$ by a factor of 2.}.
\equaref{eq:Ecore} is an analytic estimate rather than the result of a full numerical solution of the coupled non-linear field equations for the endpoint configuration in the $SU(2)\times U(1)$ model.
In the semilocal limit this expression for the monopole mass is exact in the sense that it provides the leading singularity in the $g\to 0$ limit.  More generally, there will be additional contributions on the right hand side of~\equaref{eq:Ecore} suppressed by extra powers of $g^2$. A more precise determination of such corrections is an interesting direction for future work. For now we take the monopole mass to be given by Eq.~\eqref{eq:Ecore}.

To determine the bounce action, we follow the Euclidean picture shown in the right panel of \figref{fig:zstringnucl}, reproducing the monopole–antimonopole nucleation process on a  string as described Ref.~\cite{Preskill:1992ck}.
A string sweeps out a two‑dimensional worldsheet with one timelike and one spacelike coordinate. Performing a Wick rotation converts the timelike direction into another spacelike one, converting the Lorentzian worldheet into a Euclidean surface. Quantum fluctuations may then punch a circular hole of radius \(R\) in the worldsheet and the rim of this hole is the closed worldline of the nucleated monopole–antimonopole pair.
In the thin‑wall limit, where the string thickness and the monopole core size are negligible compared with the radius of the critical bubble \(R\), the Euclidean action simplifies to
\be
S_E =
M_{\rm m} \!\oint\! ds
-\mu_{\rm str}\!\int\! d^2S \,,
\label{eq:SE}
\ee
where $M_{\rm m}$ is the monopole mass and the string tension is $\mu_{\rm str}$.
The first term of \equaref{eq:SE} describes the effect of the bubble wall tension that tries to shrink the bubble and contributes positively to the action. The second term is the two-dimensional volume effect that accounts for the energy deficit inside the bubble volume. Creating the monopole-antimonopole pair increases the action by \(2\pi R M_{\rm m}\) while removing the string patch of area \(\pi R^{2}\) lowers the action by \(\pi R^{2}\mu_{\rm str}\) resulting in the Euclidean action of the 2D bubble of the true vacuum of radius $R$ in the form
\be
S_E (R)=2\pi R\,M_{\rm m}-\pi R^{2}\mu_{\rm str}\, .
\ee
Extremising this action with respect to $R$ we find the critical bubble radius, $R=M_{\rm m}/\mu_{\rm str}$ and substituting this back into $S_E (R)$ we find the
action of the classical solution - Coleman's bounce in the thin-wall approximation:
\be
S_B=\pi\,\frac{M_{\rm m}^{2}}{\mu_{\rm str}}\,.
\ee
The result is valid as long as all scales remain much smaller than $R$.
The monopole mass in this expression corresponds to the energy contained within the monopole core, obtained in \equaref{eq:Ecore}, and combining these two equations we find, reproducing the result found in Ref.~\cite{Preskill:1992ck},
\be
\label{eq:SBub}
S_{\rm B}\,=\, \frac{8\pi^2}{g^2}\, \frac{\alpha_{\infty}\eta^2}{\mu_{\rm str}}
\,=\, \frac{8\pi^2}{g^2}\, \frac{\pi}{\alpha_{\rm str}}\,.
\ee
To ensure that the thin-defect approximation is justified, we need to check that in the desired region of the parameter space, both the monopole size,
$R_{\rm core}$ shown in \equaref{eq:Rmono}, and the string width, $\delta_s \simeq 1/M_Z$, are much smaller than the radius $R$ of the critical bubble,
\be
\label{eq:Rbub}
R \,=\, \frac{\sqrt{8\pi\, \alpha_{\infty}}}{\alpha_{\rm str}} \,\frac{1}{g \eta}\,=\, \frac{2\sqrt{2}\pi}{\alpha_{\rm str}}\,\frac{1}{g \eta}\,.
\ee
This is indeed the case,
\be
\label{eq:Thin-mono}
\frac{R_{\rm core}}{R}\,=\,
\frac{1}{2}\,\frac{\alpha_{\rm str}}{\alpha_{\infty}} \,=\,
\frac{\alpha_{\rm str}}{2\pi}
< 1\,,
\ee
and
\be
\label{eq:Thin-str-mZ}
\frac{\delta_{s}}{R}\,\simeq\, \frac{\alpha_{\rm str}}{\sqrt{2}\pi}\, \frac{g}{g'}\,\left(1+ {\cal O}(g/g')^2\right)
< 1\,,
\ee
as both of these constraints hold in the limit $g\ll g'$ and with our values of $\alpha_{\rm str}$ plotted in
\figref{fig:profilefuncs}. 
We thus conclude that the result of \equaref{eq:SBub} for the bounce action derived in the thin-defect limit is robust and the semiclassical decay rate of the metastable string takes the form
\begin{align}
 \Gamma \,\simeq\, \frac{\mu_\text{str}}{2\pi}\,e^{-S_B}
 \,=\,
\frac{\mu_\text{str}}{2\pi}\, e^{-\pi \kappa}\,,
\label{eq:decay}
\end{align}
where the quantity $\kappa$ is given by
\be
\label{eq:kappadef}
\kappa\,=\, \frac{8\pi^2}{g^2}\, \frac{1}{\alpha_{\rm str}\, (\beta)}\,,
\ee
where the dependence of $\alpha_{\rm str}$ as a function of $\beta$ is shown in the right panel of \figref{fig:profilefuncs}. 

\section{Results}
\label{sec:results}
Pulsar Timing Arrays (PTAs) are uniquely sensitive to gravitational waves (GWs) in the nanohertz frequency range. By using long observational baselines to monitor pulse arrival times from an array of millisecond pulsars, PTAs can detect a stochastic GW background (SGWB), which manifests as spatially correlated timing residuals across the sky.
In 2023, the NANOGrav collaboration reported compelling evidence for such a signal, identifying quadrupolar spatial correlations consistent with the Hellings–Downs curve~\cite{Hellings:1983fr} and strongly supporting a GW origin~\cite{NANOGrav:2023gor,NANOGrav:2023icp,NANOGrav:2023hfp,NANOGrav:2023ctt,NANOGrav:2023hvm}. This landmark discovery was independently confirmed by EPTA~\cite{EPTA:2023fyk,EPTA:2023sfo,EPTA:2023akd,EPTA:2023gyr,EPTA:2023xxk,EuropeanPulsarTimingArray:2023egv}, PPTA~\cite{Zic:2023gta,Reardon:2023zen,Reardon:2023gzh}, and CPTA~\cite{Xu:2023wog}.
While a population of inspiralling supermassive black hole binaries is the most anticipated astrophysical explanation, alternative cosmological sources remain viable. These include first-order phase transitions, cosmic superstrings~\cite{Ellis:2023tsl}, and metastable cosmic strings~\cite{Buchmuller:2023aus,Antusch:2023zjk,Madge:2023dxc}. Notably, Bayesian analyses of the PTA data already disfavour standard, stable Nambu–Goto string networks, placing strong limits on the symmetry-breaking scale for such models. This motivates a deeper investigation into alternative string scenarios. Ongoing PTA measurements, alongside future probes such as space- and ground-based interferometers and atomic sensors, offer powerful tests of grand unified theories and early-universe dynamics~\cite{Buchmuller:2013lra,Dror:2019syi,Buchmuller:2019gfy,Chigusa:2020rks,Saad:2022mzu,Fu:2023mdu}.

\begin{figure}
    \centering
    \includegraphics[width=0.75\linewidth]{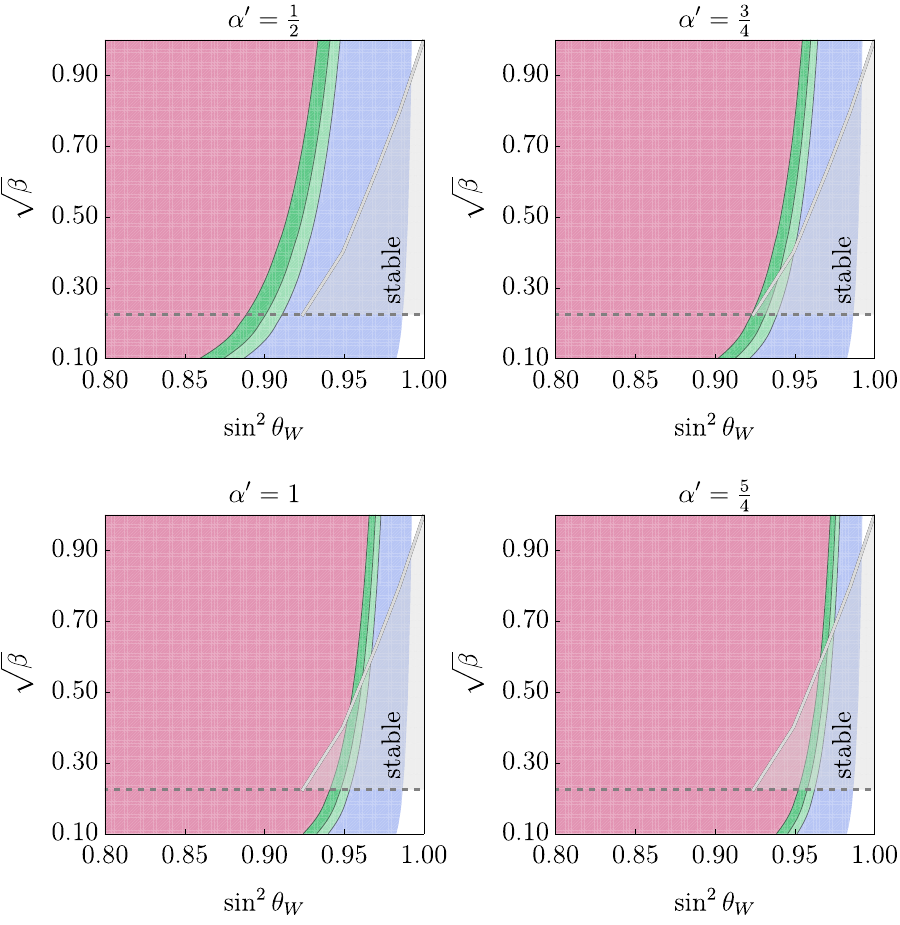}
    \caption{Regions of parameter space in the \((\beta, \sin^2\theta_w)\) plane. Shaded in pink are points with \(\kappa>8.5\); dark green corresponds to \(8.5>\kappa>8\); light green to \(8>\kappa>7.5\); and blue to \(\kappa<7.5\). The grey contour shows the metastable region {taken from Fig.~1 of Ref.~\cite{James:1992wb}, based on an analysis of the spectrum of single particle fluctuations around the classical string background.} The three panels correspond to different values  of the $U(1)$ fine structure constant of \(\alpha'=g'^2/4\pi\). An overlap between the metastable region and the \(\kappa\)-favoured band is visible for \(\alpha' \ge 3/4\), showing that the PTA-preferred range can be accommodated for moderately large \(U(1)\) gauge couplings.
}
    \label{fig:result}
\end{figure}

Our main result is summarised in \figref{fig:result}, which shows the regions of model parameter space that can consistently reproduce the PTA stochastic background~\cite{NANOGrav:2023gor, EPTA:2023fyk, Reardon:2023zen, Xu:2023wog}. If this stochastic gravitational wave background arises from the decay of cosmic strings, the signal's amplitude and spectral shape constrain the rate of string decay per unit length, $\Gamma$. This rate is, in turn, related to the parameter $\kappa$ via the semiclassical decay formula~\cite{Monin:2008mp, Monin:2009ch,Leblond:2009fq},
\begin{align}
    \Gamma \approx \frac{\mu_\text{str}}{2\pi}\, e^{-\pi\kappa}\,.
    % \label{eq:decay}
\end{align}

The NANOGrav 15-year analysis~\cite{NANOGrav:2023hvm} interprets the stochastic gravitational-wave signal in terms of two phenomenological models of metastable cosmic-string networks. The first, denoted \textsc{meta-l}, assumes that gravitational radiation is dominated by oscillating loops produced by the long-string network. The second, \textsc{meta-ls}, augments this picture by including a contribution from string segments: finite-length pieces of string bounded by monopole--antimonopole pairs. The principal physical distinction between these models lies in whether vibrational energy on the segments can efficiently radiate as gravitational waves. If unconfined gauge flux can escape from the segment endpoints, as in some microscopic realisations, then energy preferentially dissipates into gauge radiation and segment emission is suppressed, which realises the \textsc{meta-l} scenario. Conversely, if gauge flux is confined so that vibrational energy remains trapped on the segments, then segment emission can make an appreciable contribution and the behaviour corresponds to \textsc{meta-ls}. It has shown that composite monopole-string structures in $SO(10) \to SU(5) \times U(1)$ breaking patterns produce configurations where monopoles connected by strings do not carry unconfined flux after electroweak breaking~\cite{Maji:2025thf}, providing a concrete realisation that favours the \textsc{meta-ls} scenario and yields improved fits to NANOGrav 15-year data.

The analysis of~\cite{NANOGrav:2023hvm} reports that both models can reproduce the observed signal, but shows a mild preference for the \textsc{meta-ls} fit over \textsc{meta-l}, as reflected in the reported Bayes factors. Both types of behaviour arise naturally within our electroweak-like metastable string framework. In the minimal, single-scale realisation, the dark \(U(1)_\text{IR}\) remains unbroken at low energies and the magnetic flux associated with the string can escape from segment endpoints. Segment emission is therefore suppressed, which corresponds to \textsc{meta-l}. An equally straightforward variant occurs if the \(U(1)_\text{IR}\) is Higgsed at a lower scale, in which case the flux is confined and segment emission mirrors the \textsc{meta-ls} prescription. The two NANOGrav model assumptions therefore correspond to two limiting cases within the same underlying microscopic setup.

Mapping the \textsc{meta-ls} fit to our parameters yields an approximate constraint
\begin{align}
    7.5 < \sqrt{\kappa} < 8.5,
    \label{eq:kappabound}
\end{align}
which we quote as a representative range, since the \textsc{meta-l} (loops-only) fit produces a slightly narrower allowed interval that lies entirely within Eq.~\eqref{eq:kappabound}. Thus, the range in Eq.~\eqref{eq:kappabound} succinctly captures the phenomenologically viable domain for both model variants.

In the electroweak string model under consideration, the semiclassical decay rate can be computed from the properties of the string configuration. 
Our model parameter space is conveniently expressed in terms of three parameters:  $\beta,\,\sin^2\theta_w$, and the $U(1)$ gauge coupling $\alpha'$,
\be
\alpha^\prime\, = \, {g^\prime}^2/(4\pi)\,.
\ee
The value of $\kappa$ follows from Eq.~(\ref{eq:kappadef}) and the semiclassical calculation outlined in \secref{sec:bounce}. This allows us to map the \(\kappa\)-range in Eq.~(\ref{eq:kappabound}) directly onto contours in the \((\beta, \sin^2\theta_w)\) plane. 

{In \figref{fig:result}, the regions shaded light and dark green yield values for $\kappa$ inside the range indicated in Eq.~(\ref{eq:kappabound}).} The three panels of the figure show how these regions shift as a function of the parameter $\alpha^\prime$. Not all points in parameter space correspond to strings that are long-lived enough for the semiclassical decay treatment to apply. In particular, electroweak strings are only metastable in part of their parameter space. This is determined {from the stability analysis performed in Ref.~\cite{James:1992wb}, in which single particle fluctuations around the classical background given by the string profile functions are considered, and their spectrum is computed}. If all fluctuation modes have positive energy, the string configuration is metastable and can decay semiclassically with a rate governed by \(\kappa\). If, however, any fluctuation mode has negative energy, the configuration is unstable and decays promptly to particle excitations. The semiclassical description of the decay rate is not valid in this case.

The metastable region taken from the fluctuation analysis of Ref.~\cite{James:1992wb} is overlaid as a grey region in \figref{fig:result}. Only points lying inside this triangular region correspond to cosmic strings that are metastable, and can produce the gravitational wave signal under consideration. The triangular region is bounded from below by a dashed grey line. The string profile functions entering the stability analysis were only computed numerically for $\beta$ values above this bound, since this computation becomes more challenging for small $\beta$, when the two profile functions show significant variation over two different physical scales ($M_\Phi$ and $M_Z$) that are widely separated. It would be interesting to explore stability in the portion of parameter space below this line in future work.

The crucial feature of \figref{fig:result} is the overlap between the green-shaded region (corresponding to $\kappa$ values compatible with PTA observations) and the grey-shaded metastable region. This overlap occurs only for sufficiently large values of $\alpha^\prime$, as illustrated in the rightmost panel. For smaller $\alpha^\prime$, the preferred $\kappa$ region lies outside the metastable region, implying that the strings would decay too quickly to be the source of the signal. The existence of an overlap region therefore identifies a range of parameters for which metastable cosmic strings provide a viable explanation of the PTA signal.

We now reexamine the robustness of our approach for calculating the metastable string decay rate and discuss potential sources of large corrections to it.
To visualise the overlap region in~\figref{fig:result} where the PTA GW signal is reproduced we can consider a 
benchmark point,
\be
\label{eq:alprime}
\alpha' \,=\, 1\,, \quad 
\sin^2 \theta_w \,\simeq \, 0.95 \,,\quad
\sqrt{\beta} \,\simeq\, 0.3 \,,
\ee
These correspond to the $SU(2)$ gauge coupling and the Higgs self-coupling values of the order,
\be
\label{eq:al-lam}
\alpha \,=\, g^2/(4\pi)\,\simeq\, 0.05 \,,\quad 
\lambda /(4 \pi) \,\simeq\, 0.01\,,
\ee
along with the string tension values roughly around $\alpha_{\rm str} \simeq 2$
to fit the required range shown in \equaref{eq:kappabound} for the semiclassical exponent $\kappa$ in \equaref{eq:kappadef}.
With these values at hand it follows that:
\begin{enumerate}
\item the assumed near-semi-local limit $\alpha/\alpha' \ll 1$ is realised;
\item the string configuration in this region is stable classically;
\item our thin-wall approach for calculating its quantum tunnelling decay rate is justified (since the constraints Eqs.~\eqref{eq:Thin-mono}-\eqref{eq:Thin-str-mZ} are satisfied);
\item the non-Abelian $SU(2)$ sector and the Higgs sector are reasonably weakly coupled as shown in \equaref{eq:al-lam}.
\end{enumerate}
However, the $U(1)$ coupling $\alpha' \sim 1$, and one could worry if the non-weakly coupled dynamics of the Abelian sector would invalidate the semiclassical approach based on expanding the path integral around the classical string solution. We do not expect that this is the case. To concentrate on the effects of $\alpha'$  we can for simplicity decouple the non-Abelian sector. In the resulting Abelian Higgs model the string is an infinitely long-lived, localised field configuration that carries a conserved and divergenceless magnetic flux. Since no flux-free configuration can be continuously reached from the flux-carrying one, the string cannot decay into configurations where the flux is absent. Furthermore, as long as \(\alpha_{\rm str} < \alpha_{\infty} = \pi\), which holds throughout the parameter range of interest, there are no alternative flux-carrying configurations with lower energy than the string. These conclusions rely solely on flux conservation and the energetics of Abelian vortex solutions, and do not depend on whether the Abelian Higgs model is weakly or strongly coupled. The string remains a valid configuration to expand around. Corrections in $\alpha'$ will affect the form of the pre-factor in front of the exponent in the decay rate formula of \equaref{eq:decay} but we do not expect them to invalidate our conclusions. The pre-factor emerges from integrating out fluctuations around the bounce solution in the path integral and the relevant contributions are closed loops (bubble diagrams) involving propagators in the background of the bounce configurations. At one-loop level the prefactor was computed in~\cite{Monin:2008mp} which for our $SU(2) \times U(1)$ model results in,
\begin{align}
    \Gamma 
    \,=\, 
   {\rm const}\, \, \frac{\mu_\text{str}}{2\pi}\, e^{-\pi \kappa}\,,
    \label{eq:decayMV}
\end{align}
where $ {\rm const} =\frac{e}{2\sqrt{\pi}}$
and is independent of the coupling constants. Higher loop effects (two and higher loop connected bubble diagrams)
involve interaction vertices and should introduce the dependence on 
$\alpha'$ in our model, but a detailed calculation of these effects go beyond the scope of this paper. 

\section{Discussion}
\label{sec:conclusions}
In this paper we have studied metastable cosmic string configurations in a one-scale $SU(2)\times U(1)$ dark-sector model and have derived their decay rate per unit length, $\Gamma$. Starting from an electroweak-like theory with a single Higgs doublet, $\Phi$, we constructed embedded $Z$-string solutions and computed their tension by solving the Nielsen–Olesen equations numerically, thereby obtaining the dimensionless string tension coefficient $\alpha_{\rm str}(\beta)$ where $\beta = M_{\Phi}^2/M_Z^2$. In the near semi-local regime $g\ll g'$, where $g$ and $g'$ are the $SU(2)$ and $U(1)$ gauge couplings, we then evaluated the energy of the delocalised flux configuration by mapping the theory on the vacuum manifold to a non-linear sigma model, which gives the analytic result $\alpha_\infty=\pi$. Using these ingredients we computed the monopole core size and the critical bubble radius and showed explicitly that both the string width $\delta_s\sim M_Z^{-1}$ and the monopole core size $R_{\rm core}\sim M_W^{-1}$ satisfy the thin-wall criterion. This places the tunneling problem firmly in the thin-wall (thin-defect) limit, in which the Euclidean action reduces to the standard world-sheet expression and yields the semiclassical bounce action
$S_{B} \;=\; \frac{8\pi^2}{g^2}\,\frac{\alpha_{\infty}}{\alpha_{\rm str}}
          \;=\; \pi\,\kappa$.
We then mapped the resulting $\kappa$ onto the range preferred by PTA analyses of metastable string networks, and found overlapping regions of parameter space in which classically stable $Z$-strings in this one-scale model  reproduces the observed stochastic gravitational wave background.

One aspect of our general setup that we have not investigated in detail is the formation of the string network. A potential concern is whether the initial defect population is dominated by short string segments with monopole endpoints that collapse rapidly, rather than a long-lived network of ``infinite'' strings as we have assumed. However, at least in one-scale settings in the semilocal limit, it has been established that networks with significant populations of long strings do form, as reviewed in Refs.~\cite{Achucarro:1999it,Vachaspati:1984dz,Achucarro:2005vpt}.  This supports our assumption that a metastable network persists until the semiclassical decay epoch in the parameter region we consider where the non-Abelian gauge coupling $g$ is small, without the need for additional inflationary stages to dilute monopoles. This economy contrasts with many multi-stage scenarios, where an intermediate inflationary period is invoked to dilute monopoles while preserving the long-string component.

We work in a \emph{one-scale} setting, where the defect physics is controlled by a single symmetry-breaking scale, the dark Higgs vacuum expectation value, \(\eta\). This contrasts with \emph{multi-step} breaking which can also lead to the formation of metastable strings
\(SU(2)\to U(1)\to \emptyset\), in which monopoles from the first transition can nucleate on strings formed at the second~\cite{Buchmuller:2019gfy,Buchmuller:2023aus,Shifman:2002yi,Chitose:2023dam,Chitose:2025cmt}. The two gauge symmetry breaking transitions occur at the scales $V$ and $v$ respectively.
String tension is governed by the lower of the two scales, $\mu_{\rm str} \sim 2\pi v^2.$  These strings are metastable and can decay into monopole pairs by unwinding their field configurations via the $SU(2)$ gauge fields. Monopole configurations in this model are governed by the large vacuum expectation value, $M_m \sim 4\pi V/g.$ 
As shown in~\cite{Shifman:2002yi},
the resulting string decay rate can be calculated semiclassically and the approach is robust in a thin-wall limit which relies on the hierarchy of the vacuum expectation values: $V \gg v$. To fit the semiclassical rate exponent $\kappa$ in these models into the range relevant for the PTA signal, one is forced however to forgo the large hierarchy of scales and consider the regime $V \to v$ where the two steps of the symmetry breaking pattern are not well-separated~\cite{Buchmuller:2023aus}. In this case, it is not a priori clear that the corresponding string configuration is classically stable in the regime $V \sim v $; if it is not, the string would undergo rapid classical decay, overwhelming the exponentially suppressed semiclassical decay process. These features of two-scale models are different from the single-scale model setup considered in our paper where the string configuration itself is known to be classically stable, and its quantum mechanical decay rate can be reliably calculated.

\bibliographystyle{JHEP}
\bibliography{ref}

\end{document}